# Direct measurement of the localized-itinerant transition, hybridization and antiferromagnetic transition of 5f electrons

D. H. Xie[1], M. L. Li[2], W. Zhang[1], L. Huang[1], W. Feng[1], Y. Fang[1], Y. Zhang[1], Q. Y. Chen[1], X. G. Zhu[1], Q. Liu[1], B. K.Yuan[1], L. Z. Luo[1], P. Zhang[2], X. C. Lai[1#] and S. Y. Tan[1*]

[1]*Science and Technology on Surface Physics and Chemistry Laboratory, Mianyang 621908, China*

[2]*Institute of Applied Physics and Computational Mathematics, Beijing 100088, China*

[*]*To whom correspondence should be addressed:* tanshiyong@caep.cn *and* laixinchun@caep.cn

**In heavy-fermion compounds, f electrons show both itinerant and localized behaviours depending on the external conditions, and the hybridization between localized f electrons and itinerant conduction bands gives rise to their exotic properties like heavy-fermions, magnetic orders and unconventional superconductivity. Due to the risk of handling radioactive actinide materials, the direct experimental evidences of the band structure evolution across the localized-itinerant and magnetic transitions for 5f electrons remain lacking. Here, by using angle-resolved photoelectron spectroscopy, we revealed the dual nature (localized and itinerant) and the development of two different kinds of heavy quasi-particle bands of 5f electrons in antiferromagnetic (AFM) $USb_2$. Partially opened BCS-type energy gaps were observed on one quasi-particle 5f band below the AFM transition around 203 K, indicating that the magnetic orders in $USb_2$ are of spin density wave (SDW) type. The localized 5f electrons and itinerant conduction bands hybridize to form another heavy quasi-particle band around 120 K, and then open hybridization gaps at even lower temperature. Our results provide direct spectral demonstration of the localized-itinerant transition, hybridization and SDW transition of 5f electrons for uranium-based materials.**

# I. INTRODUCTION

Uranium-based materials display intriguing and attractive properties, such as heavy-fermion states, unconventional superconductivities, and multiple orderings[1-8]. It is generally believed that these properties mainly originate from the interplay between partially filled shell of 5f orbitals and a very broad band of conduction electrons. The 5f electrons have an intermediate character between localized 4f electrons of rare-earth compounds and itinerant 3d electrons of transition metals, the wide variety of physical properties of uranium-based materials may stem from the dual character of 5f electrons. The f electrons behave as atomic local moments at high temperatures, and the local moments combine with the conduction electrons (c-f hybridization) to form a fluid of very heavy quasi-particles (QP) as the temperature is decreased[9]. It is a great challenge to understand how the itinerant low-energy excitations emerge from the localized moments, which requires the understanding of how the dual characters of the 5f electrons manifest themselves in the band structures and physical properties of the uranium-based materials.

Another hotly debated issue in uranium-based materials is the origin of the magnetic order. Generally, in rare-earth 4f compounds, a long-range magnetic order can be understood by the Ruderman-Kittel-Kasuya-Yoshida (RKKY) interaction mechanism, which is essentially based on a localized f -electron picture. On the other hand, the origin of magnetism in actinide 5f compounds has not been well understood since the 5f electrons show magnetic properties of both itinerant and localized character[10]. There are very few limited cases where the origin of magnetism has been directly revealed for uranium based materials. For example, itinerant ferromagnetism was demonstrated by photoemission experiments in UTe[11]. However, the direct experimental evidence of the band structure evolution across the antiferromagnetic transition for uranium based materials remains lacking.

Antiferromagnetic $USb_2$ provides an ideal platform to study the dual nature (localized and itinerant) of 5f electrons and the mechanism of antiferromagnetic transition for uranium-based materials, as it is a moderately correlated electron system with a quasi-2D electronic structure[12-16]. Angle-resolved photoemission spectroscopy (ARPES) is a powerful method for studying the U 5f electronic states. However, the extremely small energy scale of the uranium-based materials has made it difficult to be studied in detail. The most commonly used soft x-rays ARPES[17] strongly enhances the bulk part of the 5f signal, but the much poorer energy resolution prevents the detection of its fine structures, in particular close to $E_F$. Previous ARPES studies[18-22] observed a narrow heavy QP band below $E_F$ and the first kink structure of actinide materials in $USb_2$. However, only the normal emission ARPES spectrum was reported, the complete electronic structure and the band structure evolution across the antiferromagnetic transition are lacking. Furthermore, an ultrafast optical spectroscopy study[23] revealed the opening of multiple gaps in $USb_2$, indicating the existence of c-f hybridization and magnon-mediated band renormalization.

In this paper, we report the Fermi surface topology and complete band structures of $USb_2$ by ARPES using 21.2 eV light source. While the 5f bands around Γ and M show weak dispersion and itinerant character, we observe two non-dispersive bands at -20 meV and -60 meV below $E_F$ over the entire Brillouin zone, indicating that the 5f electron bands are partially localized and partially itinerant in $USb_2$. The localized 5f electrons and itinerant conduction bands hybridize to form one heavy quasi-particles around 120 K, and then open hybridization gaps at lower temperature. Possible Fermi surface nesting condition and partially opened energy gaps were observed on another heavy 5f electron bands of $USb_2$, indicating that the magnetic orders in $USb_2$ are of SDW type.

## II. EXPERIMENTAL METHODS

**Sample synthesis and transport measurement**. The high-quality single crystals of $USb_2$ were grown from Sb flux with a starting composition of U:Sb = 1:15. X-ray diffraction measurements were performed on a PANalytical X'Pert Pro diffractometer (Cu K$_\alpha$-radiation) from 10 ° to 90 ° with a scanning rate of 6 ° per minute. The dc magnetic susceptibility in fields of 10 kOe applied along the a- and c-axes in the temperature range 2–300 K was performed using a commercial PPMS-9 system (Quantum Design). Electrical resistance at temperatures of 2-300 K was measured by a four-point ac method. The coefficient of electronic specific heat determined by specific heat measurement in the temperature range of 2–10 K is 25 mJ.K$^{-2}$. mol$^{-1}$.

**ARPES measurement.** The samples were cleaved in-situ along the (001) plane and measured under ultrahigh vacuum better than 5 x 10$^{-11}$ mbar. The in-house ARPES measurements were performed with SPECS UVLS discharge lamp (21.2 eV He-I$\alpha$ light). All data were collected with Scienta R4000 electron analyzers. The overall energy resolution was 15 meV or better, and the typical angular resolution was 0.2 °. The lowest temperature of our system is 10 K using liquid helium. A freshly evaporated gold sample in electrical contact with the USb$_2$ sample was served to calibrate $E_F$.

**LDA calcualtion.** The first-principles calculation for the band structure of USb$_2$ was performed with the project augmented wave method as implemented in the *Vienna Ab-initio Software Package* (VASP). The plane waves less than the energy of 600 eV are used to expand the wave functions. The local spin-density approximation (LSDA) exchange-correlation functional was adopted. U 5f, 6s, 6p, 6d and 7s electrons and Sb 5s and 5p electrons are treated as valence electrons. The spin-orbit coupling effect was included throughout the calculations. The internal positions of the atoms were relaxed until the residual forces on all relaxed atoms were smaller than 0.01eV/ Å, while the lattice parameters were fixed at the experimental lattice parameters of a=4.27 Å and c=8.748 Å. We have simulated the low-temperature antiferromagnetic phase as shown in Fig. 1(a) with a unit cell of 12 atoms. The first Brillouin zone was sampled in the k-space with Monkhorst-Pack scheme and the grid size was 11×11×5.

# III. RESULTS

**Sample characterization and band structures.** Uranium dipnictides $USb_2$ crystallizes in the tetragonal structure of anti-$Cu_2Sb$ type ($D^7_{4h}$ or P4/nmm), which orders antiferromagnetically below a high Néel temperature of 203 K[24,25]. Magnetic moments of U ions are aligned ferromagnetically in the (001) planes, which are stacked along the [001] direction in an antiferromagnetic sequence (↑↓↓↑) in $USb_2$. The residual resistivity ratio (RRR=$\rho_{RT}/\rho_{2K}$) of $USb_2$ single crystal is 150, which is much higher than the previous reported results of 80 in the literature[13,24]. The temperature dependence of magnetic susceptibility of $USb_2$, measured along a and c axis, display sharp maximums confirming the Néel temperature of about 203 K (Fig.1c). The measured electrical resistivity curves along a and c directions show clear kinks across the antiferromagnetic transition (Fig.1d). The low-energy electron diffraction (LEED) pattern of $USb_2$ shows typical tetragonal symmetry without noticeable lattice or charge superstructures, as shown in Fig.1e.

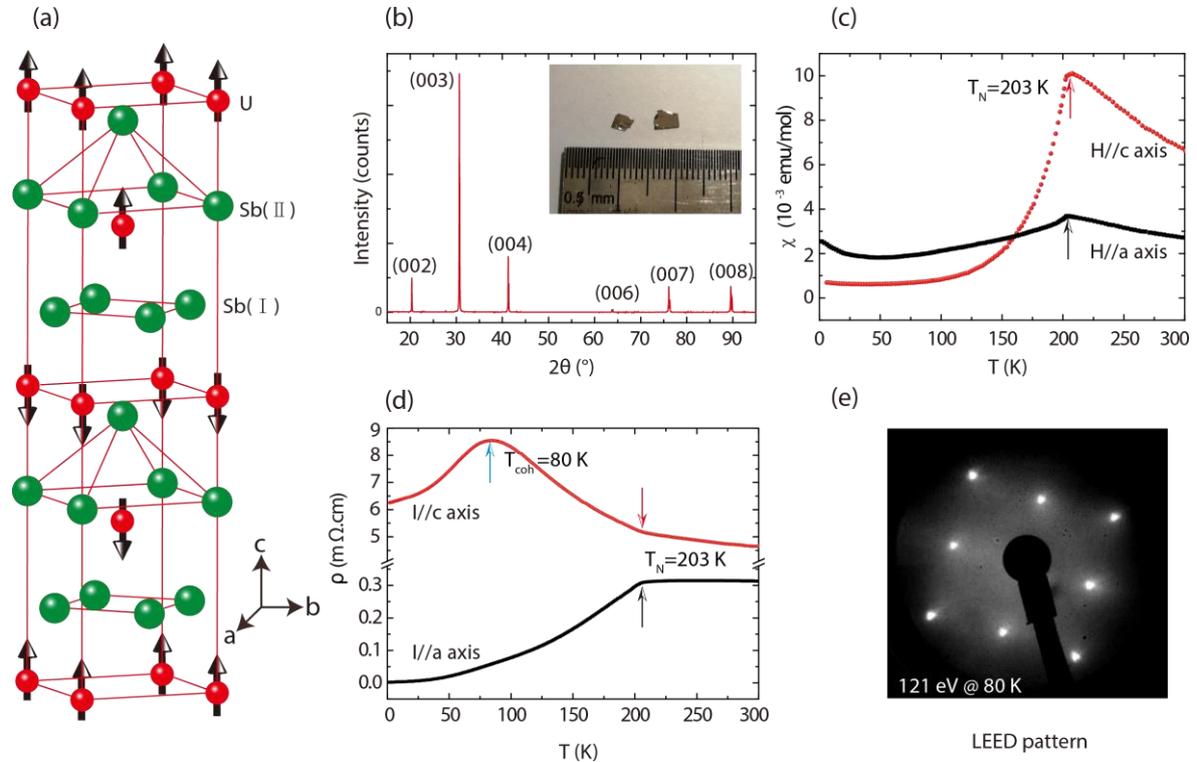

**Fig.1** The crystal structure and transport properties of $USb_2$. (a) The crystallographic and magnetic structure of $USb_2$, arrows indicate the direction of the ordered magnetic moments along c axis. (b) X-ray

diffraction pattern and picture of an USb$_2$ single crystal. (c) The temperature dependence of magnetic susceptibility along a and c axis, T$_N$ stands for the Néel temperature. (d) The temperature dependence of electrical resistivity along a and c axis, T$_{coh}$ stands for the coherence temperature. (e) The low-energy electron diffraction (LEED) pattern measured at 80 K.

The Fermi surface topology and overall band structures are shown in Fig.2. The observed Fermi surface consists of two elliptical electron pockets (β and β′) around each X point, a diamond hole pocket (η) around Γ point, as shown in Fig.2(a). Furthermore, there are large area spectra intensities around Γ and M points, which are originated from U 5f bands. The photoemission intensities measured along Γ–X and Γ–M directions over a large energy scale are presented in Fig.2(b). Lots of highly dispersive bands originated from itinerant conduction electrons (U 6d or Sb 5p electrons) exist at high binding energy below -0.3 eV. Two extremely sharp photoemission peaks can be observed in the vicinity of $E_F$, which are mainly originated from U 5f electrons, as marked in the integrated energy distribution curves (EDCs) in Fig.2(b3).

Figs. 2(c) and 2(d) show the low energy band structure of USb$_2$ along the high-symmetry directions. A nearly flat band named α in the vicinity of $E_F$ can be observed around Γ point along both Γ–X and Γ–M directions in Figs.2(c1) and 2(d1), while a η band with very steep energy dispersion (large Fermi velocity) can be clearly seen along Γ–M directions. Below α band, a broad and weak band named ε can be observed. There are two narrow bands γ and δ (each one is composed of two nearly degenerate bands) located below $E_F$ around M point, as shown in Figs.2(d2). Two electron-like bands β and β' cross $E_F$ and form the elliptical pockets around X. The observed valence bands and low energy bands qualitatively agree with our first principle calculations with local spin-density approximation (LSDA) as shown in Fig.2(e) and Fig.S1(Supplementary material). The renormalization factors differ from each other for different bands, it is estimated to be about 2 and 6 for the conduction band (β/β') and the flat 5f band (α/γ), respectively.

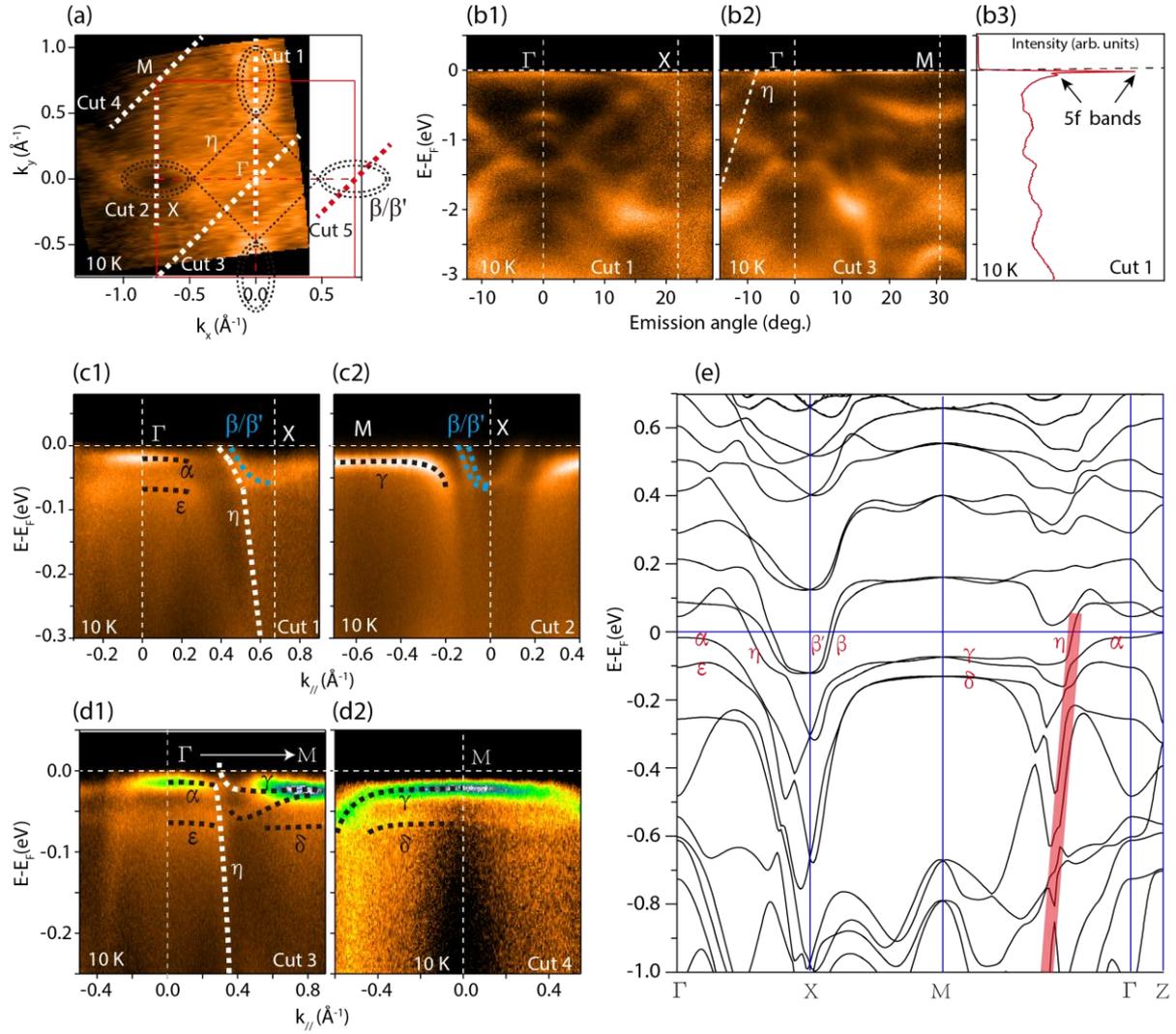

**Fig.2** The electronic structure of $USb_2$. (a) Fermi surface topologies of $USb_2$ measured at 10 K integrated over a [$E_F$ -10 meV, $E_F$ +10 meV] window. (b1-b2) The valence band structure along Γ-X and Γ-M directions. (b3) The integrated photoemission spectra along Γ-M direction. (c1-c2) The low-energy band structure of $USb_2$ around Γ and X points. (d1-d2) The low-energy band structure of $USb_2$ around Γ and M points. (e) The calculated band structure of $USb_2$ by local spin-density approximation (LSDA). The red solid line indicates the location of fast dispersion band η as experimental observed in (b2).

Our calculated band structure is in reasonably good accord with two previous theoretical studies[26,27]. However, there is still some minor discrepancy between different calculated band structures at the near $E_F$ region. The exact positions of the band top (bottom) for α, γ bands (β/β′ bands) differ with each other, which might due to the fact that different methods and approximations are adopted. For all the calculations, the 5f electrons were treated as itinerant and no correlation correction was considered. Due to the absent of correlation effects, none of the calculations fully agree with experimental, especially at the near $E_F$ region.

Early de Haas-van Alphen (dHvA) results[13,24] and DFT calculation[26] found four Fermi surface contours in USb$_2$, two circular electron pockets around Γ point and two circular hole pockets around X point, as shown in Fig.S4(b). Later, Yang[27] and Durakiewicz[21] reported only three Fermi surfaces(one hole pocket around Γ, and two electron pocket around X) according to their ARPES and theoretical simulation. The fourth hole pocket was renormalized below $E_F$ at Γ, which is consistent with our results. The cross section of the Fermi surface contour from our ARPES and the dHvA results[13] are shown in Fig.S4, which show good accordance. In one Brillouin zone, there is one η hole pocket (occupies 16.2% of the Brillouin zone), two β electron pockets(2×5.4%=10.8%), two β' electron pockets (2×2.8%=5.6%). The hole Fermi surface thus occupy 16.2% and the electron Fermi surface occupy 16.4%, which shows that USb$_2$ is a compensated metal with equal carrier number of electrons and holes.

The α and γ bands around Γ and M points exhibit weak dispersion and angular dependence, which are originated from the U 5f orbitals with itinerant nature in USb$_2$.These flat bands can no longer be seen in ThSb$_2$, which can be seen as the f$^0$ reference system shown in Fig. S3. Interestingly, we observed two straight-line shaped bands (non-dispersive bands with localized nature) at about -20 meV and -60 meV below $E_F$, as shown in Figs. 3(a) and 3(b). The intensity of the -20 and -60 meV straight line shaped bands is very weak compared with other bands, they can hardly been seen along the high symmetry direction when high intensity quasi-particle bands exist. We choose cut 5 in a special direction as shown in Fig.2(a) far away from the high intensity quasi-particle bands(ie.α and γ), so that the weak intensity -20 and -60 meV bands can be seen more clearly. It is noteworthy that the -20 and -60 meV straight line shaped bands does not correspond to any of the bands in Fig. 2(e), they can't be reproduced in either our LDA calculation or the previously published calculations[26,27]. The ARPES observed -20 and -60 meV straight line shaped bands qualitatively agree with the two crystalline electric field (CEF) excited states based on a crystal field model calculation[28], in which the splitting of the two lowest CEF excited states from ground state is estimated to be 262 cm$^{-1}$ (32 meV) and 428 cm$^{-1}$ (53 meV)[28]. Our ARPES results show that the 5f electron bands are partially localized and partially itinerant in USb$_2$.

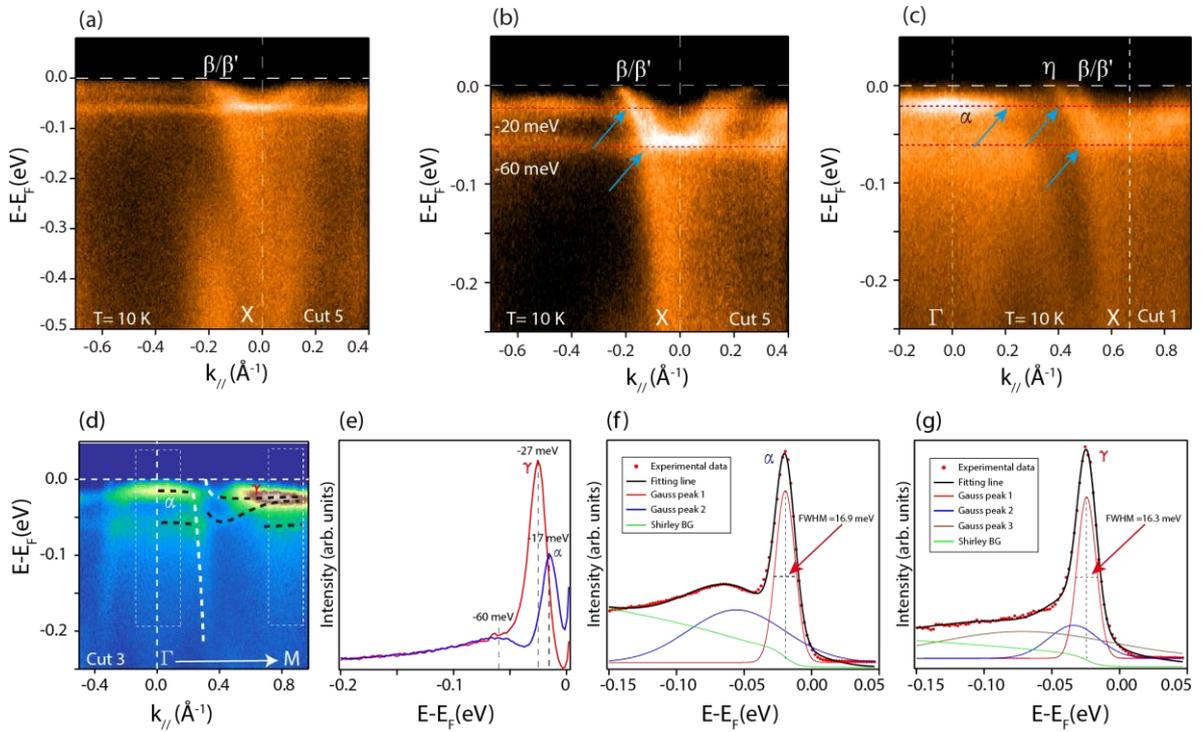

**Fig.3** The electronic structure of USb$_2$. (a) The low-energy band structure of USb$_2$ around X point, the location of cut 5 is shown in Fig.2(a). (b) The blown-up near E$_F$ region of (a). (c) The low-energy band structure of USb$_2$ along Γ-X direction，blue arrows mark kinks in the dispersion. (d) The low-energy band structure of USb$_2$ along Γ-M direction. (e) The quasi-particle peaks of α and γ bands, the EDCs are integrated over the white rectangle area in (d) after division by the Fermi-Dirac function. (f) The Gauss Fit of the EDC around Γ point, two Gauss peaks and a Shirley type background were used to fit the experimental data. (g) The Gauss Fit of the EDC around M point, three Gauss peaks and a Shirley type

background were used to fit the experimental data.

Strong interactions in correlated electron systems may result in the formation of heavy quasiparticles that exhibit kinks in their dispersion relation. The first kink structure in f-electron systems was observed in USb$_2$ on α quasi-particle band at Γ point[21]. Kinks can also be found on η and β conduction bands with energy scale of -20 and -60 meV, as shown in Figs.3(b) and 3(c). Kinks are taken as evidence for the coupling of electrons to either phonons or spin fluctuations in cuprates[29,30]. While for the antiferromagnetic compound USb$_2$, the mechanism of the multiple kink structures and its energy scales is a mystery right now. Further theoretical interpretation is called to unveil the underlying physics of the kink structures found in USb$_2$.

Extremely narrow 5f-derived band around Γ point has been reported in the literature[20], whose full width at half-maximum (FWHM) of the 5f peak is about 24 meV. Two quasi-particle bands named α and γ can be clearly resolved in Fig.3 (d), while the former correspond to the one reported in literature. The FWHM of the 5f bands around Γ and M points are estimated to be 16.9 meV and 16.3 meV, respectively. Although α and γ bands both exhibit narrow but dispersive 5f band feature, these two quasi-particle bands differ from each other in many ways, as shown in Fig.3(e). Firstly, the photoemission intensity of γ band is always stronger than that of α band. Secondly, the α and γ bands both show gaped feature at 10 K, but the gap size of γ band (∼27 meV) is much bigger than that of α band (∼17 meV). Thirdly, the α and γ bands both exhibit sharp quasi-particle peaks, but the line-shapes have essential difference, indicating their different origins.

**Localized-itinerant transition and c-f hybridization of 5f electrons.** We conducted detailed temperature dependent measurements on the narrow 5f band around Γ point to reveal the mechanism of the heavy QP. In Fig. 4(a8), we can see only one conduction band named η which cross $E_F$ at 130 K, the 5f electrons behave like atomic local moment at such high temperature at Γ point. With decreasing temperature, the atomic 5f electrons begin to hybridize with conduction band (η band), form heavy QP band (α band) and open hybridization gaps at low temperature, as shown in Fig.4(a1). The band structure around Γ point after division by the Fermi-Dirac function at 11 K and 130 K are shown in Fig.4(b). We can see that heavy QP band emerges after c-f hybridization, and open hybridization gaps at 11 K. Fig.4(c) show the illustration of the c-f hybridization, all the bands are extracted from Fig.4(b).

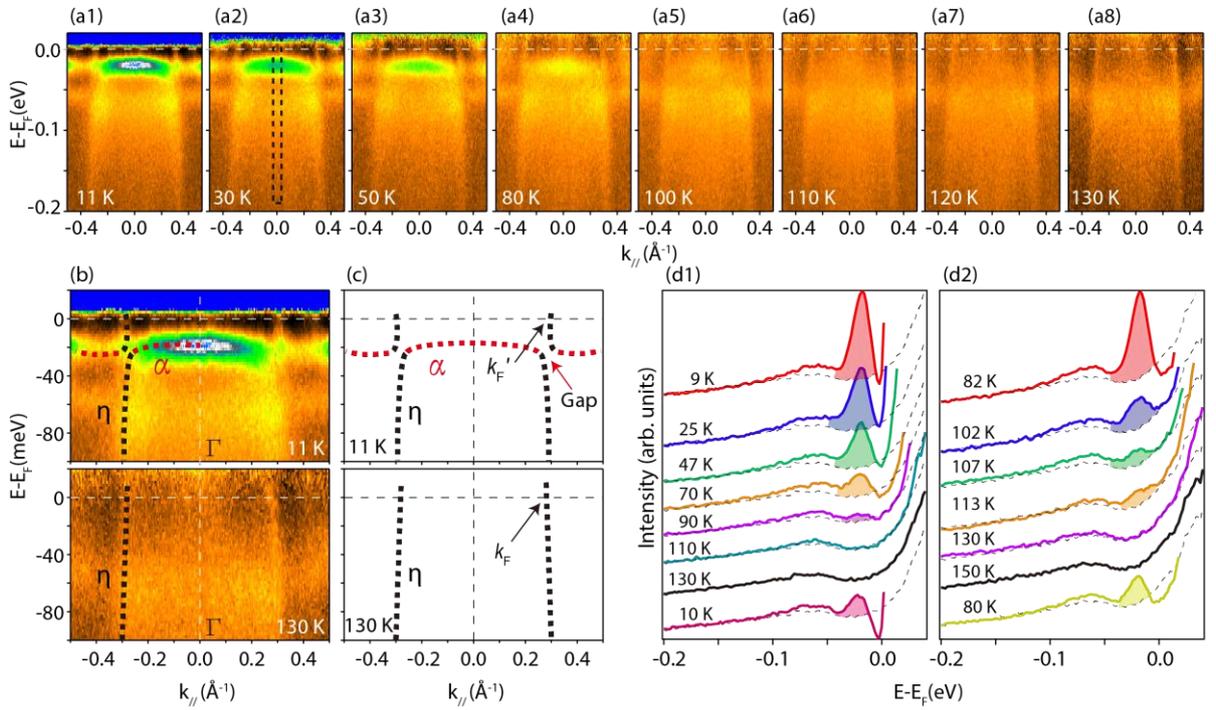

**Fig.4** The development of quasi-particle band around Γ point in USb$_2$. (a) Temperature dependence of the band structure around Γ point, taken from sample 1#. The spectrums are divided by a Fermi-Dirac function convolved with a Gaussian function representing the instrumental resolution to remove the thermal broadening contribution[31], as illustrated in Fig.S5 in the supplementary material. (b) The band structure around Γ point after division by the Fermi-Dirac function at 11 K and 130 K, respectively. (c) The schematic drawing of the band structure extracted from (b). (d) Temperature dependence of the EDCs at Γ point after division by the Fermi-Dirac function, the EDCs are integrated over the black rectangle area as illustrated in (a2).The temperature is varied from 9 to 130 K and cooled town to 10 K again in (d1) for sample 2#, the dashed line correspond to the spectrum at 130 K, and the shaded regions represent the difference from the 130 K data; while it is varied from 82 to 150 K and cooled town to 80 K again in (d2) for sample 3# , the dashed line correspond to the spectrum at 150 K, and the shaded regions represent the difference from the 150 K data.

The electronic specific heat of $USb_2$ of 25 mJK$^{-2}$ mol$^{-1}$ is small in comparison with Ce-based heavy fermions, but is typical for many uranium compounds where strong hybridization between 5f and conduction electrons leads to itinerant behavior. In Ce-based heavy fermions, the flat quasi-particle bands formed by strong c-f hybridization usually cross $E_F$, thus give very big effective masses in the order of 100-1000 $m_e$. In $USb_2$, the heavy 5f quasi-particle band around Γ was pushed away from $E_F$ without Fermi crossing (thus has no contribution to the effective masses), which is mainly caused by the magnetic order. If the heavy 5f quasi-particle band was not gaped and cross $E_F$, it should give a much heavier effective masses. That is the case[8] for $UPd_2Al_3$, heavy quasi-particle band was observed which cross $E_F$ in its paramagnetic state with a big electronic specific heat of 210 mJK$^{-2}$ mol$^{-1}$.

We take the EDCs at Γ point to determine the onset temperature of the QP band development more precisely. The shaded regions in Figs.4(d1) and (d2) represent the spectrum weight of heavy QP band α, the data are taken from two samples. In Fig.4(d1) taken from sample 2#[The temperature dependence of the band structure is shown in Fig. S6], the QP peaks are very strong at 9 K, but the intensity becomes weaker as temperature increases and vanishes at about 110 K. When the sample is cooled down back to 10 K, the QP peak appears again but the intensity is much weaker than that of the initial state. The data shown in Fig.4(d1) are reproducible with non-negligible sample surface degradation during a low-high-low temperature cycle. To minimize the aging effect, sample 3# is measured from 82 K which is more close to the transition temperature. Again, sharp QP peak can be found at 82 K as shown in Fig.4(d2), and the QP intensity becomes weaker with increased temperature and vanishes at about 130K. Considering the sample degradation over time, the onset temperature of the QP band development is estimated to be about 120 ±10 K. The $k_F$ and $k_F'$ in Fig.4(c) represent the Fermi crossing of the η band at 130 K and 11 K, respectively. The size of $k_F'$ is larger than $k_F$, as shown in Fig. S7, indicating that the Fermi surface volume of η band gets larger at low temperature after the localized-itinerant transition.

The localized-itinerant transition of the f electrons and the emerging of coherence peak with a characteristic temperature $T^*$ have been the central issue of heavy electron physics. Kotliar[9] et al. have used the single-site LDA+DMFT method to simulate the localized-itinerant transition of 4f electrons in $CeIrIn_5$. At room temperature, there is very little spectral weight at the Fermi level because the f electrons are tightly bound and localized on the Ce atom. As the temperature is decreased, a narrow coherence peak appears near the Fermi level, and the area of the peak can be interpreted as the degree of f electron delocalization. Mo et al. performed detailed temperature dependent ARPES measurements[32] on heavy fermion $YbRh_2Si_2$ and revealed the development of QP states, in the form of sharp, weakly dispersive peaks, below a characteristic temperature($T^*\sim 50$ K) higher than its coherence temperature ($T_{coh}\sim 25$ K). The direct observation of how the heavy fermion state develops in $CeCoIn_5$ was reported from two independent group recently[33,34]. Our ARPES observed c-f hybridization, localized to itinerant transition and the development of coherence peaks ($T^*\sim 120$ K) of 5f electrons in $USb_2$ agree well with the theoretical predication and experimental finding of 4f electrons, indicating that the 4f and 5f electrons may share the same underlying physics.

**Spin density waves transition of 5f electrons.** The detailed temperature dependence of the narrow 5f QP band around M point is presented in Fig. 5. At 82 K in Fig.5(a1), which is well below the antiferromagnetic phase transition temperature 203 K, the intensity of narrow γ band is very strong and it lies below $E_F$ (with an energy gap of 24 meV). At 210 K in Fig.5(a7), which is above the phase transition temperature, the intensity of γ band gets weaker due to the losing of coherence and the band crosses $E_F$. We tracked the symmetrized EDCs on the γ band around M point to reveal the gap opening behavior more precisely. At 210 K above the transition temperature, there is large density of states at $E_F$ [Fig. 5(b)]. The density of states at $E_F$ is obviously suppressed with decreasing temperature, an energy gap opens at 195 K below the phase transition temperature of 203 K. The gap size increased with decreasing temperature, saturated at low temperature. The largest gap size is about 27 meV at 10 K, which give a ratio of $2\Delta/k_BT_N \sim 4$. Fig.S9 in the supplementary materials presents the temperature dependence of the band structure along X-M direction, which shows the similar temperature dependent behavior.

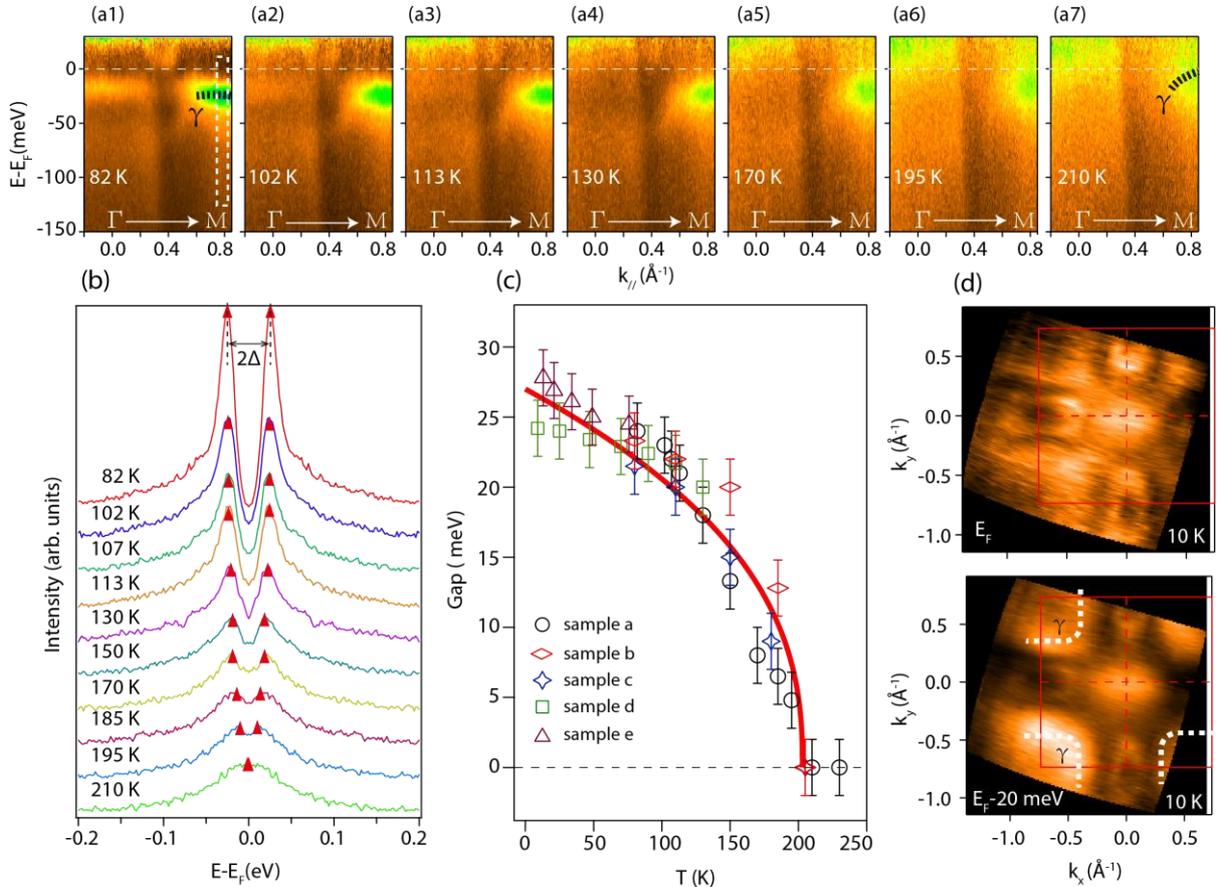

**Fig.5** The spin density wave transition around M point in USb$_2$. (a) Temperature dependence of the band

structure along Γ-M direction. (b) Temperature dependence of the symmetrized EDCs around M point, the EDCs are integrated over the white rectangle area in (a1). The distance between the two red triangles correspond to 2Δ at each temperature. (c) Temperature dependence of the SDW gap. The gap data were collected from five different samples, and a constant erro bar of ±2 meV was adopted. The solid line is the fit to a mean field formula: $\Delta=\Delta_0 (1-T/T_N)^\psi$, where $\Delta_0$=27 meV, $T_N$=203 K, $\psi$=0.4 ± 0.1. (d) Constant-energy maps at different binding energies of $E_F$ and $E_F$ - 20meV, respectively.

The surface degradation over the duration of the experiment (aging effect) and ultra high transition temperature (the temperature should be ranged from the lowest (10 K) to higher than 203 K) make it a difficult task to get valid temperature dependent data in one sample. We plot the temperature dependent energy gaps from five different samples in Fig.5(c), and used a mean field formula $\Delta=\Delta_0 (1-T/T_N)^\psi$ to fit the experimental data[35]. The method of determination of the energy gap is shown in Fig.S8, a constant error bar of ±2 meV was adopted. The experimental data qualitatively agree with the BCS formula with the exponent $\psi$=0.4 ± 0.1. The slight deviation of the experimental data from the mean field formula may due to the strong correlation of 5f electrons and the c-f hybridization at low temperature.

ARPES has been one of the most powerful tools to investigate the electronic structure renormalization cross the magnetic transition. For localized spin antiferromagnetic transition, the potential difference induced by localized moment ordering is very small that the band electrons can hardly feel the potential change, resulting in negligible or non-detectable changes in the photoemission signals[36]. For itinerant spin antiferromagnetic transition, band folding/splitting[37,38], band gapping[35], band shifting/large scale spectrum weight transfer[39] are all characteristic signatures which have been observed in literatures. However, Fermi surface nesting, partial gapping of the Fermi surface and the BCS-like temperature dependence of the gap size are only observed for SDW type antiferromagnetism[35], which can be recognized as the fingerprint of

SDW.

In the conventional picture of spin density wave transition[40], the formation of electron-hole pairs with a nesting wave vector connecting parallel regions of FSs would lead to the opening of an energy gap. As the SDW gap is observed on the γ band[the white dashed lines shown in Fig.5(d)], propagation vector connecting the parallel Fermi surface sheets between different γ bands may exist in the $k_x$-$k_y$ plane or along the $k_z$ direction. To identify this, photon energy dependent ARPES study utilizing synchrotron radiation source is needed to obtain the three dimensional Fermi surface map of $USb_2$ in the future. Although we can't pin down the exact nesting vector at present, our ARPES observed partially gapping of the Fermi surface and the BCS-like temperature dependence of the gap size are very strong evidences of SDW in $USb_2$. In addition, an ultrafast optical spectroscopy study[23] has also revealed that $USb_2$ opens an energy gap below $T_N$ with BCS-like T dependence: $\Delta_S \approx 46.1(1-T/T_N)^{0.5}$ (meV), supporting that the magnetic order in $USb_2$ is of SDW type.

Although there are a number of studies on the magnetism of actinide-based compounds, there are only a few cases where the origin of magnetism has been directly revealed, especially the electronic structure evolution cross the AFM transition have not been well understood due to the lack of experimental electronic structure studies. Our ARPES results on $USb_2$ provide the first spectrum demonstration of the electronic structure evolution of uranium based materials across the SDW transition. Opening of an energy gap found in $USb_2$ indicates that the SDW transition behavior of 5f electron materials is similar to $Cr$[35], while band folding and splitting are often found in the parent compound of iron-based superconductors across the SDW transition[37,38].

## V. DISCUSSION AND CONCLUSION

Heavy electron materials are often described as a Kondo lattice that is composed of an array of interacting local moments of 4f or 5f electrons coupled antiferromagnetically to a conduction electron sea. There are many phenomenological models[41-43] to explain the complex properties of heavy Fermion compounds. The key of these existing theoretical models involves a transfer of the f-electron spectral weight from the local moment component to the itinerant heavy electrons with decreasing temperature, and $T_{coh}$ is the coherence temperature marking the onset of the process. It is of fundamental and practical importance to learn what determines the magnitude of $T_{coh}$, and what happens about the electronic structure and transport properties cross $T_{coh}$.

The increase of $\rho_c$ in USb$_2$ below $T_N$ was firstly interpreted to be the enhanced quasi-two-dimensionality of the Fermi surface as a consequence of the doubling of the elementary cell along the c direction[13]. According to our ARPES results, we think that the combination of SDW ordering and Kondo screening may be responsible for the large hum of $\rho_c$ in USb$_2$, and the resistivity maximum stands for its coherence temperature. The formation of energy gaps and the decrease of the effective number of conduction electrons on the γ band cause the resistivity rise below $T_N$, and the Kondo scattering further enhance this resistivity rise. With decreased temperature, the buildup of phase-coherent intersite coupling in the spin-flip scattering causes the resistivity downturn. The temperature scale for the development of the quasi-particle band α around Γ point matches the temperature of the resistivity peak.

In most heavy fermions, heavy electrons caused by the hybridization between local moments and conduction electrons emerge gradually as temperature falls below $T_{coh}$, the competing between the Kondo effect and the RKKY interaction determine the ground state. If the Kondo interaction dominates, a Fermi liquid state may then be stabilized at a lower temperature. On the contrary, the ground state is a magnetically ordered state ($T_N < T_{coh}$). It is unique in USb$_2$ that it magnetically orders at rather high $T_N$=203 K, and the development of coherence by c-f hybridization at $T_{coh} < T_N$. The Kondo screening and magnetic interaction coexist and compete at the ground state, which give strong constraint on the existing theoretical model. The same phenomenon has also been found[44] in the an antiferromagnetic Kondo lattice $CeRh_2Si_2$ [$T_N \approx$ 38 K, the third highest among Ce systems]. They observed strong interaction between valence and f electrons at a temperature well below $T_N$, which demonstrates the importance of hybridization effects in the antiferromagnetic phases of Kondo lattices.

In $USb_2$, the 5f electrons are partially localized and partially itinerant. When the temperature is reduced, the itinerant parts of 5f electrons form SDW states around 203 K and develop heavy QP band around M point. When the temperature is further decreased, some of the localized 5f electrons hybridize with conduction bands to form another heavy electron band around Γ point at T*(∼120±10K). The $T_{coh}$ is estimated to be about 80 K by the peak position of the electrical resistivity along c axis[Fig.1d], and the onset temperature (T*) of the localized-itinerant transition of $USb_2$ is determined to be 120±10K, indicating that the localized-itinerant transition sets in at a temperature much higher than $T_{coh}$. It is worthwhile to note that the Kondo interaction and magnetical interaction happen on different 5f bands in the momentum space, and they coexist and interact with each other at the ground state.

In summary, we revealed the dual nature (localized and itinerant) and the development of two different kinds of heavy fermions in antiferromagnetic $USb_2$. We present clear spectrum evidence of the localized-itinerant transition and the c-f hybridization of 5f electrons for uranium based heavy-fermion materials. The localized-itinerant transition is found to sets in at a temperature much higher than $T_{coh}$, and the Fermi surface become larger after the localized-itinerant transition. Furthermore, partially opened energy gaps were observed on one heavy quasi-particle 5f bands, giving the first spectrum demonstration how the band structure evolves cross the SDW transition for uranium based itinerant antiferromagnets. Our results provide clear microscopic picture of how the heavy electrons develop and evolve with temperature in uranium based materials, which is important for a deep understanding of exotic properties in heavy-fermion compounds.

# ACKNOWLEDGMENTS

We gratefully acknowledge helpful discussions with Prof. D. L. Feng, H. Q. Yuan, X. Dai, Y. F. Yang and J. B. Qi. This work is supported by the Science Challenge Project (Grants No. TZ2016004), the National Science Foundation of China (Grants No. 11504341, 11504342, U1630248) and the National Key R&D Program of China.